\documentclass[aps,prd,tightenlines,twocolumn,floatfix]{revtex4}
\usepackage{graphicx}
\usepackage{epsfig}
\usepackage{dcolumn}
\usepackage{amsmath}
\usepackage{bm}

\def\ep{\epsilon}

\def\ra{\rangle}
\def\be{\begin{equation}}
\def\ee{\end{equation}}
\def\bea{\begin{eqnarray}}
\def\eea{\end{eqnarray}}

\begin{document}

\title{Zero-Mode Contribution in
Nucleon-Delta Transition}

\author{Jianghao Yu}
\affiliation{Department of Physics, Peking University, Beijing
100871, China}
\author{Teng Wang}
\affiliation{Department of Physics, Peking University, Beijing
100871, China}
\author{Chueng-Ryong Ji}
\affiliation{Department of Physics, North Carolina State University,
Raleigh, NC 27695-8202, USA}
\author{Bo-Qiang Ma}
\affiliation{Department of Physics, Peking University, Beijing
100871, China} \affiliation{MOE Key Laboratory of Heavy Ion Physics,
Peking University, Beijing 100871, China}

\begin{abstract}
We investigate the transition form factors between nucleon and
$\Delta$(1232) particles by using a covariant
quark-spectator-diquark field theory model in (3+1) dimensions.
Performing a light-front calculation in parallel with the manifestly
covariant calculation in light-front helicity basis, we examine the
light-front zero-mode contribution to the helicity components of
light-front good~(``+") current matrix elements. Choosing the
light-front gauge ($\epsilon^+_{h=\pm}=0$) with circular
polarization in Drell-Yan-West frame, we find that only the helicity
components $({1\over 2}, {1\over 2})$ and $({1\over 2},-{1\over 2})$
of the good current receive the zero-mode contribution. Taking into
account the zero-mode, we find the prescription independence in
obtaining the light-front solution of form factors from any three
helicity matrix elements with smeared light-front wavefunctions. The
angular condition, which guarantees the full covariance of different
schemes, is recovered.
\end{abstract}

\pacs{12.38.Bx, 13.40.Gp, 14.20.Dh, 14.20.Gk}

\maketitle

\renewcommand{\baselinestretch}{1.2}
\normalsize

\section{Introduction}
\label{sec.1}

The study of nucleon transition form factors has long been
recognized as one of the essential steps towards a deep
understanding of strong interaction. The transition process between
nucleon and $\Delta$(1232) resonance, the lowest baryon excitation,
is important to investigate the internal quark and gluon structure
of the nucleon and its resonance. When hadronic systems are
described in terms of quarks and gluons, it is part of nature that
the characteristic momenta are of the same order or even very much
larger than that of the masses of the particles involved. Thus, a
relativistic treatment is one of the essential ingredients for a
successful model description. The light-front constituent quark
model (LFQM) within the framework of light-front~(LF)
dynamics~\cite{Bro98} appears to be a useful phenomenological tool
which keeps the relativistic effect into account while it
incorporates some basic spin-flavor structure of the hadron.

In the formalism of LFQM, the good~(``+") component of current
matrix elements for the analysis of the hadron form factors can be
divided into two parts. One is the valence part, in which the ``+"
current matrix elements are diagonal in the Fock state expansion,
that can be expressed in terms of the convolutions of LF
wavefunctions.  The other is the nonvalence part, in which the ``+"
components are only related to the off-diagonal elements in the Fock
state expansion. In the $q^+\to 0$ limit, where $q^+$ is the
longitudinal component of the momentum transfer $q$ ($q^\pm=q^0\pm
q^3$, $q^2=q^+q^--{\vec q}^2_\perp$), it was usually taken for
granted that the nonvalence part vanishes and thus one only needs to
take into account the valence part or the diagonal elements in the
Fock state expansion. This is true in the cases when the hadrons
involved are spin-0 bosons and spin-1/2 fermions as in most previous
investigations. The ``+" component of matrix elements is therefore
denoted as ``good" component. However, in recent analyses of spin-1
form factors~\cite{BCJ02,BCJ03,CJ04,CJ05}, it has been shown that
the nonvalence part has non-vanishing contribution in the $q^+\to 0$
limit, which is called {\it zero mode}. Without the zero mode
contribution, the angular condition~\cite{GK}, which assures that
different ways of extracting the form factors produce the same
physical results, would be violated.

In this paper we analyze the zero mode contribution to the
transition form factors in the $N$-$\Delta$ transition process. As
our purpose is for a recovery of full covariance in the LFQM
formalism by including the zero-mode contribution in the
$N$-$\Delta$ process but not to present a realistic model which can
fit the available data, we adopt a simple but exactly solvable
quark-spectator-diquark model for the sake of simplicity and
clarification. We specify the kinematics in Drell-Yan-West~(DYW)
frame with the LF gauge ($\ep^+_{h=\pm}=0)$, which has been used in
the LFQM analysis. To investigate the zero mode contribution in a
clear way, we make the taxonomical decomposition of the full current
matrix elements into valence and nonvalence contributions.
Performing calculation in the LF helicity basis we find that indeed
the good component of the currents has non-vanishing zero mode
contribution in this model. In particular, there is zero mode
contribution in the LF helicity amplitudes $(h',h)=({1\over 2},
{1\over 2})$ and $({1\over 2},-{1\over 2})$, where $h$ and $h'$ are
the initial and final helicities, respectively.  When the number of
current matrix elements gets larger than the number of physical form
factors, the conditions that the current matrix elements must
satisfy are essential to test the underlying theoretical model for
the hadrons. The analysis of $N$-$\Delta$ transition process
requires in general eight light-front helicity amplitudes although
there are only three physical form factors. Thus, there must be five
conditions for the helicity amplitudes. Using the light-front parity
relation, one can reduce the number of helicity amplitudes down to
four. One more relation among the remaining four helicity amplitudes
comes from the conservation of angular momentum and this relation is
called {\it angular condition}. Consequently, only three helicity
amplitudes are independent of each other as it should be because
there are only three physical form factors involved in the
$N$-$\Delta$ transition process. However, there are several
different prescriptions~\cite{Weber90,Simula95} in choosing
independent current matrix elements to extract the three physical
form factors. We take four prescriptions as examples, and show that
different types of helicity combinations produce the same results
and the angular condition is satisfied if the zero mode contribution
is included in each description. We do this taking the baryon as a
system of quark and diquark. Note that there are different ways of
treating the diquarks in the literature. One way is called the
quark-spectator-diquark model in which one would re-organize the
spectators by an effective diquark to take into account the
flavor-spin structure of the whole spectator when any of the quark
is struck. In this framework, there is no struck diquark and one
should have an additional factor of 3 to account for that there are
three valence quarks in the baryon. Another way is called the
quark-diquark model in which the diquarks are treated as independent
particles and one should also take into account the contribution
when diquarks are struck. However, the zero-mode issue is completely
independent from whether the model is ``spectator diquark" or
``struck diquark". That is to say that the Lorentz covariance must
work independently for each contribution. In this work, we thus
compute only one part of the full contribution neglecting the flavor
structure and show that the covariance is nevertheless recovered on
its own with the inclusion of the zero-mode contribution. The struck
diquark contribution would independently work essentially in the
same way.

This paper is organized as follows.  In Section~\ref{sec.2}, we
present the Lorentz-invariant transition form factors and angular
conditions in LF helicity basis. The four prescriptions used in
extracting the physical form factors are also briefly discussed. In
Section~\ref{sec.3}, after taking into account the Melosh
transformation~\cite{Melosh} for instant-form SU(6) quark-diquark
wave functions, we get effective covariant vertices for the nucleon
and
$\Delta$ coupling with the quark and the diquark. 
In Section~\ref{sec.4}, we present our model calculations in both
the manifestly covariant Feynman method and the LF technique. In the
$q^+=0$ frame, we separate the full amplitudes to the valence and
nonvalence parts to determine zero mode contribution by a power
counting method. In Section~\ref{sec.5}, we present the numerical
results of transition form factors and angular condition.
Conclusions are given in Section~\ref{sec.6}. In Appendix, we
summarize our results of the LF helicity amplitudes for each
helicity component used in our calculation.

\section{Transition form factor in light-front helicity basis}
\label{sec.2}

The $N$-$\Delta$ transition matrix element of the electromagnetic
current $J^\mu$ between the initial $|p,h\ra$ and the final
$|p',h'\ra$ states can be written as
\begin{equation}
J_{h'h}^{\mu}=\left\langle\, p',h'\,\left| J^\mu
\right|\,p,h\,\right\rangle
=\bar{u}_\rho(p',h')\,\Gamma^{\rho\mu}\,u(p,h) , \label{eqn.1}
\end{equation}
where $u(p,h)$ is the nucleon Dirac spinor with momentum $p$ and
helicity $h$,  and $u_\nu(p',h')$ is the Rarita-Schwinger
spin-${3\over 2}$ spinor for the $\Delta$ particle with momentum
$p'$ and helicity $h'$, respectively. Here the covariant tensor
$\Gamma^{\rho\mu}$ is defined by a linear combination of three
kinematic-singularity-free transition form factors $G_1(Q^2)$,
$G_2(Q^2)$ and $G_3(Q^2)$:
\begin{eqnarray}
\Gamma^{\mu\nu} &=& G_{1}(Q^{2})\,\Big(q^{\mu}\gamma^{\nu}-\gamma
\cdot q \,g^{\mu\nu}\Big)\,\gamma_5
\nonumber\\
&&+G_{2}(Q^2)\,\Big(q^{\mu}\,{(p+p')^{\nu}\over 2}-{(p+p')\cdot
q\over 2}\, g^{\mu\nu}\Big)\,\gamma_5 \nonumber\\
&&+G_{3}(Q^2)\,\Big(q^{\mu}\,q^{\nu}-q^{2}\,g^{\mu\nu}\Big)\,\gamma_5
,\label{eqn.2}
\end{eqnarray}
where  $q^\mu=(p'-p)^\mu$ is the four-momentum
transfer and $Q^2=-q^2$.

The physical Sachs-like form factors with definite multipole or
helicity, such as magnetic dipole $G_M(Q^2)$, electric quadrupole
$G_E(Q^2)$ and Coulomb quadrupole $G_C(Q^2)$ form factors, are
related in a well-known way to the three kinematic-singularity-free
transition form factors~\cite{Jones,DEK} as follows
\begin{widetext}
\begin{eqnarray}
G_{M}(Q^{2})&=&
\frac{M_{P}}{3(M_{P}+M_{\Delta})}\Big[{(3M_{\Delta}+M_{P})(M_{\Delta}+M_{P})+Q^{2}\over
M_{\Delta}}\,G_{1}(Q^{2})+(M_{\Delta}^{2}-M_{P}^{2})\,G_{2}(Q^{2})-2Q^{2}\,G_{3}(Q^{2})\Big]
,\nonumber\\
G_{E}(Q^{2})&=&
\frac{M_{P}}{3(M_{P}+M_{\Delta})}\Big[{M_{\Delta}^{2}-M_{P}^{2}-Q^{2}\over
M_{\Delta}}\,G_{1}(Q^{2})+(M_{\Delta}^{2}-M_{P}^{2})\,G_{2}(Q^{2})-2Q^{2}\,G_{3}(Q^{2})\Big]
,\nonumber\\
G_{C}(Q^{2})&=&
\frac{M_{P}}{3(M_{\Delta}+M_{P})}\Big[4M_{\Delta}\,G_{1}(Q^{2})
+(3M_{\Delta}^{2}+M_{P}^{2}+Q^{2})\,G_{2}(Q^{2})+2(M_{\Delta}^{2}-M_{P}^{2}-Q^{2})\,G_{3}(Q^{2})\Big]
,\label{eqn.3}
\end{eqnarray}
\end{widetext}
where $M_{\Delta}$ and $M_{P}$ are the masses of
$\Delta$(1232) and nucleon.

The reference frame is taken as the Drell-Yan-West (DYW)
frame~\cite{DYW} where $q^{+}=0$ and
$q^2=q^+q^--\vec{q}^2_{\perp}<0$. In this frame, the momenta of
the nucleon and $\Delta$(1232) particle are assigned as
\begin{eqnarray}
q^{\mu}&=&
\left(0,\frac{\vec{q}_{\perp}^2+M_{\Delta}^2-M_{P}^2}{p^+},
\vec{q}_{\perp}\right), \nonumber \\
p^{\mu}&=&
\left(p^+,\frac{M_{P}^2}{p^+}, \vec{0}_{\perp}\right),\nonumber \\
p'^{\mu}&=&\left(p^+,{\vec{q}_{\perp}^2+M_{\Delta}^2\over
p^+},\vec{q}_{\perp}\right).\label{eqn.4}
\end{eqnarray}
Here, we use the notation $p=(p^+,p^-,p^1,p^2)$ and the metric
convention $p^2=p^+p^- -{\vec p}^2_\perp$ with $p^\pm=p^0\pm p^3$.

For the spin-${3\over 2}$ particle such as $\Delta$(1232), in the LF
formalism the Rarita-Schwinger spinor is
\begin{eqnarray}
u^{\mu}(p^+,\vec{p}_\perp,\frac{3}{2})&=&\epsilon^{\mu}\,
(p^+,\vec{p}_\perp,+1)\,u(p^+,\vec{p}_\perp,\uparrow),
\nonumber\\
u^{\mu}(p^+,\vec{p}_\perp,\frac{1}{2})&=&\sqrt{\frac{2}{3}}\,
\epsilon^{\mu}(p^+,\vec{p}_\perp,0)\,u(p^+,\vec{p}_\perp,\uparrow)
\nonumber\\
&&+\sqrt{\frac{1}{3}}\,\epsilon^{\mu}(p^+,\vec{p}_\perp,+1)\,
u(p^+,\vec{p}_\perp,\downarrow). \label{eqn.5}
\end{eqnarray}
Following Bjorken-Drell convention, $u(p^+,\vec{p}_\perp,h)$ is the
light-front spinor with light-front momentum $(p^+,\vec{p}_\perp)$
and helicity $h$ and $\epsilon^{\mu}(p^+,\vec{p}_\perp,h)$ is the
transverse($h=\pm$) or longitudinal($h=0$) polarization vector
with the convention $\varepsilon^\mu = (\varepsilon^+,
\varepsilon^-, \varepsilon^1, \varepsilon^2)$. With $\ep^+(p,\pm)=0$
from the LF gauge, the general form of the LF polarization vector
is given by
\begin{eqnarray}
\ep^\mu(p,\vec{p}_\perp,\pm)&=&\biggl(0,
\frac{2{\vec\ep}_\perp(\pm)\cdot{\vec
p}_\perp}{p^+}, {\vec\ep}_\perp(\pm)\biggr), \nonumber\\
\epsilon ^{\mu}(p^+,\vec{p}_\perp,0)&=&{1\over M_{\Delta}}
\left(p^+,{\vec{p}_{\perp}^2-M_{\Delta}^2\over p^{+}},p^1
,p^2\right),\label{eqn.6}
\end{eqnarray}
which satisfies $p\cdot\ep(\pm)=0$ with ${\vec\ep}_\perp(\pm)=\mp
{1\over \sqrt{2}}(1,\pm i)$.

In DYW frame, the covariant form factors can be determined using
only the plus component of the current, $J^+_{h'h}(0)\equiv\langle
P',h'|J^+|P,h\rangle$. As one can see from Eq.~(\ref{eqn.2}), eight
matrix elements of $J^+_{h'h}$ can be assigned to the current
operator, while there are only three independent invariant form
factors. However, according to LF discrete symmetry, the current
matrix elements $J^+_{h'h}$ must be constrained by the invariance
under the LF parity~\cite{Ang}
\begin{equation}
J_{-h', -h}^+=\eta^\prime_{_\Delta} \eta_{_P}
(-1)^{S^\prime-S+h^\prime-h}\,J_{h^\prime ,h}^+ ,
\label{eqn.7}
\end{equation}
where $\eta^\prime_{_\Delta}$ and $\eta_{_P}$ are the intrinsic
parity of $\Delta$(1232) with spin $S'$ and nucleon with spin $S$.
Since the initial and final states are different in $N$-$\Delta$
transition, the LF time-reversal condition does not reduce the
number of matrix elements but requires $J^+_{h'h}$ to be real.
Therefore, one can reduce the eight matrix elements of the plus
current $J_{h^\prime ,h}^+$ down to four using the LF helicity
basis, and we take $J^+_{{3\over 2},{1\over 2}}$, $J^+_{{3\over
2},-{1\over 2}}$, $J^+_{{1\over 2},{1\over 2}}$ and $J^+_{{1\over
2},-{1\over 2}}$. The plus components of current matrix elements are
related to the kinematic-singularity-free form factors $G_{1,2,3}$
with the following four relations
\begin{eqnarray}
  \begin{array}{lll}
    J^+_{\frac{3}{2},\frac{1}{2}}&=& \sqrt{2}\,q^L\Big[2G_1 +
(M_{\Delta} -M_{P}) G_2 \Big], \\
    J^+_{\frac{1}{2},\frac{1}{2}} &=& \sqrt{{2\over 3}}{Q^2\over
M_{\Delta}}\Big[-2 G_1 -(2M_{\Delta} -M_{P})G_2\\
& &+2 (M_{\Delta} -M_{P}) G_3 \Big], \\
    J^+_{\frac{1}{2},-\frac{1}{2}} & =&\sqrt{{2\over 3}}{q^L\over
M_{\Delta}}\Big[2M_{P} G_1 - 2Q^2 G_3\\
& & + (Q^2-(M_{\Delta} -M_{P})M_{\Delta}) G_2
\Big], \\
    J^+_{\frac{3}{2},-\frac{1}{2}} & =&-\sqrt{2}(q^L)^2G_2.
  \end{array}\label{eqn.8}
\end{eqnarray}
 From above relations, it is rather obvious that the four helicity
components are not independent because there are only three physical
form factors to be extracted. Thus, there must be an additional
constraint on the current matrix elements. Indeed, the rotation
invariance of the system (angular momentum conservation) offers an
additional constraint on the current operator, which yields the so
called ``angular condition" $\Delta(Q^2)$~\cite{Ang} given by
\begin{eqnarray}
\Delta(Q^2)&=&q^L[Q^2-M_{P}(M_{\Delta}-M_{P})]J^+_{\frac{3}{2},\frac{1}{2}}
\nonumber\\
&& +\sqrt{3}(q^L)^2M_{\Delta} J^+_{\frac{1}{2},\frac{1}{2}}
\nonumber\\
&& +\sqrt{3}q^LM_{\Delta}(M_{\Delta}-M_{P})J^+_{\frac{1}{2},-\frac{1}{2}}\nonumber\\
&&-[(M_{\Delta}-M_{P})(M_{\Delta}^2-M_{P}^2)+M_{P}Q^2]J^+_{\frac{3}{2},-\frac{1}{2}}\nonumber\\
&=&0.\label{eqn.9}
\end{eqnarray}

Because of the angular condition, only three helicity amplitudes are
independent as expected. However, the relations between the physical
form factors and the matrix elements $J^+_{h'h}$ are not uniquely
determined because the number of helicity amplitudes is larger than
that of physical form factors. So one still has the freedom of
choice in extracting the form factors by using different matrix
elements. For example, Weber~\cite{Weber90} and Cardarelli {\it et
al.}~\cite{Simula95} used helicity components $(h',h)=({3\over
2},{1\over 2}), ({1\over 2},{1\over 2})$ and (${1\over 2},-{1\over
2}$), which is denoted as ``Scheme A" for this prescription. In this
scheme, the kinematic-singularity-free transition form factors in
terms of the matrix elements $J^+_{h'h}$ are given by
\begin{eqnarray}
G_1^A&=&{M_{\Delta}\over q^L
K_A}\left[\left((M_{\Delta}-M_{P})^2+Q^2\right)J^+_{\frac{3}{2},\frac{1}{2}}\right.\nonumber\\
&&\left.+ \sqrt{3}(M_{\Delta}-M_{P})^2J^+_{\frac{1}{2},-\frac{1}{2}}
\right.\nonumber\\
&&\left.+\sqrt{3}q^L(M_{\Delta}-M_{P})J^+_{\frac{1}{2},\frac{1}{2}}\right],
 \nonumber \\
G_2^A&=&{2\over q^L K_A
}\left[\left((M_{\Delta}-M_{P})M_{P}-Q^2\right)J^+_{\frac{3}{2},\frac{1}{2}}
\right.\nonumber\\
&&\left.-\sqrt{3}(M_{\Delta}-M_{P})M_{\Delta}
J^+_{\frac{1}{2},-\frac{1}{2}}\right.\nonumber\\
&&\left.-\sqrt{3}M_{\Delta}
q^L J^+_{\frac{1}{2},\frac{1}{2}}\right],\nonumber \\
G_3^A&=&{1\over q^L K_A}\left[\left(M_{\Delta}^2+M_\Delta
M_P-M_P^2-Q^2\right) J^+_{\frac{3}{2},\frac{1}{2}}\right.\nonumber\\
&&\left.+\sqrt{3}\,q^L\,M_\Delta\left(M_\Delta^2-M_P^2-Q^2
\right)J^+_{\frac{1}{2},\frac{1}{2}}/Q^2
\right.\nonumber\\
&&\left.- \sqrt{3}M_\Delta^2J^+_{\frac{1}{2},-\frac{1}{2}}\right],
\label{eqn.10}
\end{eqnarray}
where
$K_A=2\sqrt{2}(M_{\Delta}-M_{P})(M_{\Delta}^2-M_{P}^2)+2\sqrt{2}M_{P}Q^2$.
Keeping the (${1\over 2},{1\over 2}$) component that gives the most
dominant contribution in the high momentum region, we can use
helicity $({3\over 2},{1\over 2}), ({1\over 2},{1\over 2})$, and
(${3\over 2},-{1\over 2}$) amplitudes to determine covariant form
factors, which we call ``Scheme B", as follows
\begin{eqnarray}
G_1^{B}&=&\frac{1}{2\sqrt{2}q^L}
J^+_{\frac{3}{2},\frac{1}{2}}+\frac{M_{\Delta}
-M_{P}}{2\sqrt{2}(q^L)^2} J^+_{\frac{3}{2},-\frac{1}{2}},
 \nonumber \\
G_2^{B}&=&-\frac{1}{\sqrt{2}(q^L)^2} J^+_{\frac{3}{2},-\frac{1}{2}}, \nonumber \\
G_3^{B} &=&{1\over 2\sqrt{2}(M_\Delta-M_P)}\Big[ {1\over
q^L}J^+_{\frac{3}{2},\frac{1}{2}}+{\sqrt{3}M_\Delta\over
Q^2}J^+_{\frac{1}{2},\frac{1}{2}}\nonumber\\
&&- \frac{M_\Delta}{(q^L)^2}
J^+_{\frac{3}{2},-\frac{1}{2}}\Big].\label{eqn.11}
\end{eqnarray}
Alternatively we can avoid using the helicity $({1\over 2},{1\over
2})$ but use helicity $({3\over 2},{1\over 2}), ({1\over 2},-{1\over
2})$, and (${3\over 2},-{1\over 2}$) amplitudes~(``Scheme C")
\begin{eqnarray}
G_1^{C}&=&\frac{1}{2\sqrt{2}q^L}
J^+_{\frac{3}{2},\frac{1}{2}}+\frac{M_{\Delta}
-M_{P}}{2\sqrt{2}(q^L)^2} J^+_{\frac{3}{2},-\frac{1}{2}},
 \nonumber \\
G_2^{C}&=&-\frac{1}{\sqrt{2}(q^L)^2} J^+_{\frac{3}{2},-\frac{1}{2}}, \nonumber \\
G_3^{C} &=&{1\over 2\sqrt{2}Q^2}\Big[ {M_P\over
q^L}J^+_{\frac{3}{2},\frac{1}{2}}-{\sqrt{3}M_\Delta\over
q^L}J^+_{\frac{1}{2},-\frac{1}{2}}\nonumber\\
&&+\frac{M_\Delta^2-M_P^2-Q^2}{(q^L)^2}
J^+_{\frac{3}{2},-\frac{1}{2}}\Big].\label{eqn.12}
\end{eqnarray}
 From above schemes, we find that the only difference between
``Scheme B" and ``Scheme C" is  the  expression of $G_3(Q^2)$. If we
take $G_3^{B} =G_3^{C}$, we can recover the angular condition given
by Eq.~(\ref{eqn.9}). In the above schemes A, B and C, the helicity
$({3\over 2},-{1\over 2})$, $({1\over 2},-{1\over 2})$ and $({1\over
2},{1\over 2})$ are avoided in Eqs.~(\ref{eqn.10}), (\ref{eqn.11})
and (\ref{eqn.12}), respectively. Since there is one more helicity
$({3\over 2},{1\over 2})$ that can be avoided, it is natural to
consider another scheme using helicity components $({3\over
2},-{1\over 2}), ({1\over 2},{1\over 2})$, and (${1\over 2},-{1\over
2}$) and get ``Scheme D":
\begin{eqnarray}
G_{1}^D &=&\frac{\sqrt{3}M_{\Delta
}}{K_D}\Big[J_{\frac{1}{2},\frac{1}{2}}^{+}+\frac{ (M_{\Delta
}-M_{N})}{q^{L}}J_{\frac{1}{2},-\frac{1}{2}
}^{+}\nonumber\\
&&-\frac{(M_{\Delta
}-M_{N})^{2}+Q^{2}}{\sqrt{3}(q^{L})^{2}}J_{\frac{3}{
2},-\frac{1}{2}}^{+}\Big],\nonumber\\
G_{2}^D
&=&-\frac{\sqrt{2}}{2(q^{L})^{2}}J_{\frac{3}{2},-\frac{1}{2}}^{+},
\nonumber \\
G_{3}^D &=&\frac{\sqrt{3}}{K_D}\Big[\frac{M_{\Delta }M_{N}}{Q^{2}}J_{\frac{1}{2},\frac{1}{%
2}}^{+}+\frac{M_{\Delta }}{q^{L}}J_{\frac{1}{2},-\frac{1}{2}%
}^{+}\nonumber\\
&&+\frac{M_{\Delta }^{2}+Q^{2}-M_{\Delta }^{2}-M_{\Delta }M_{N}
}{\sqrt{3}(q^{L})^{2}}J_{\frac{3}{2},-\frac{1}{2}}^{+}\Big],\label{eqn.13}
\end{eqnarray}
where $K_D=2\left[(M_{\Delta }-M_{N})M_{N}-Q^{2}\right]$. If we take
$G^A_2=G^D_2$ in Eqs.~(\ref{eqn.11}) and~(\ref{eqn.13}), then we
also get the ``angular condition" $\Delta(Q^2)$. Of course, we also
have other choices that involve all four helicity amplitudes in
various ways. We will show later that  different schemes are
equivalent and give the same physical result by taking into account
the zero mode contribution.

\section{Quark-spectator-diquark model description}
\label{sec.3}

In light-front dynamics, the set of LF wavefunctions provides a
frame-independent description of hadrons in Fock state expansion. In
the light-front quark model, the LF wavefunction of a composite
system can be obtained by transforming ordinary equal-time
(instant-form) wavefunction in the rest frame into LF wavefunction.
This can be done by taking into account relativistic effects such as
the Melosh rotation~\cite{Melosh}. For the lowest excited baryon
such as proton and $\Delta$(1232), a convenient Fock state basis is
the quark-diquark two-body state, which has been adopted in
Refs.~\cite{Ma96,Bro02,Qing02}.

The proton spin wavefunction in the SU(6) quark-diquark model in the
instant-form can be written as~\cite{Diquark76}
\begin{eqnarray}
\Psi^{\pm{1\over 2}}_P(qD)&=&\pm\frac{1}{3}
\left[V^0(ud)u^{\uparrow,\downarrow}-
\sqrt{2}V^{\pm1}(ud)u^{\downarrow,\uparrow}\right.\nonumber\\
&&\quad\quad\left.-\sqrt{2}V^0(uu)d^{\uparrow,\downarrow}+
2V^{\pm1}(uu)d^{\downarrow,\uparrow}\right]\nonumber\\
&&+S(ud)u^{\uparrow,\downarrow},\label{eqn.14}
\end{eqnarray}
where $\uparrow,\downarrow$ label the spin projection
$J_q^z$=$\pm\frac{1}{2}$ of the quark, $V^{s_z}(q_1q_2)$ stands for
the $q_1q_2$ axial vector diquark with the third spin component
$s_z$, and $S(ud)$ stands for a $ud$ scalar diquark. In the
inclusive and exclusive processes involving above quark-diquark
wavefunction, the diquark serves as an effective spectator to
account for the spin-flavor structure of whole spectators and do not
serve as a struck particle to contribute to the electro-magnetic
properties~\cite{Diquark76,Ma96,Qing02}. This is to say, when any of
the quarks is struck, we just take it as the quark and the
spectators are treated as an effective diquark, and we call this
kind of description as the quark-spectator-diquark model.
Alternatively, when one takes another kind of proton SU(6) spin
wavefunction such as in Ref.~\cite{Rolnick}, the diquark is treated
as an independent particle which can be a struck particle. In such a
model, one would not get the right electro-magnetic properties in
exclusive processes unless the struck diquark contribution is
included. In this work we aim at the generic structure of zero mode
contribution and take into account only the quark-spectator-diquark
contribution neglecting the flavor structure in the above two kinds
of instant-form wave functions.

Without the flavor structure, the proton SU(6) spin wavefunctions in
Eq.~(\ref{eqn.14}) and Ref.~\cite{Rolnick} can be expressed
as~\cite{KW,Weber98}
\begin{eqnarray}
\Psi^{S}_P(qD)&=&\sum_{s_1,s_2,s_3}(\chi^\dagger_{s_1}\vec{\sigma}{\rm
i}\sigma_2\chi^*_{s_2})\cdot(\chi^\dagger_{s_3}\vec{\sigma}\chi_{_S})V^{s_1+s_2}(q_1q_2)q_3^{s_3}\nonumber\\
&& +\sum_{s_1,s_2,s_3}(\chi^\dagger_{s_1}{\rm
i}\sigma_2\chi^*_{s_2})(\chi^\dagger_{s_3}\chi_{_S})S^{0}(q_1q_2)q_3^{s_3},\label{eqn.15}
\end{eqnarray}
where $\chi_s$ is the instant-form Dirac spinor with helicity $s$ in
the rest frame, and $V^{s_1+s_2}(q_1q_2)q_3^{s_3}$ and
$S^{0}(q_1q_2)q_3^{s_3}$ are given in the instant-form spin basis.
We transform the instant spinor $\chi_s$ into the LF Dirac spinor
$u_{_{LF}}(\lambda)$ with LF helicity $\lambda$ by the Melosh
transformation~\cite{Melosh}
\begin{eqnarray}
\chi_s=u_{_{LF}}(\lambda)R_{s,\lambda},\label{eqn.16}
\end{eqnarray}
where
\begin{eqnarray}
R_{s,\lambda}={1\over \sqrt{2k^+(k^0+m)}}\left(%
\begin{array}{cc}
  k^++m & k^1-{\rm i}k^2 \\
  -k^1-{\rm i}k^2 & k^++m \\
\end{array}%
\right). \label{eqn.17}
\end{eqnarray}
After the Melosh transformation, the proton spin wavefunction in LF
helicity basis is given by
\begin{eqnarray}
\Psi^{h}_P(qD)_{_{LF}}&=&(\bar{u}_1\gamma^\mu
\vec{\tau}G\bar{u}_2^{T})\cdot(\bar{u}_3\gamma_{\mu}\gamma_5\vec{\tau}
u_P)V^{\lambda_1+\lambda_2}(q_1q_2)q_3^{\lambda_3}\nonumber\\
&&+(\bar{u}_1\gamma_5G\bar{u}_2^{T})(\bar{u}_3
u_P)S^{0}(q_1q_2)q_3^{\lambda_3},\label{eqn.18}
\end{eqnarray}
where $G={\rm i}\tau_2C=G^T$ is the $G$ parity with the
charge-conjugation operator $C={\rm i}\gamma^2\gamma^0$ and
$u_i=u_i(\lambda)$ are the LF spinors. We now recognize
$\bar{u}_1\gamma^\mu \vec{\tau}G\bar{u}_2^{T}$ as the axial vector
diquark $\varepsilon^{\mu}_{_V}$ and
$\bar{u}_1\gamma_5G\bar{u}_2^{T}$ as the scalar diquark $\phi_{_S}$.
Thus, we can rewrite the LF proton spin wavefunctions in
Eq.~(\ref{eqn.18}), modulo isospin terms, as
\begin{eqnarray}
\vert\,P,h\,
\rangle&=&\bar{u}(\lambda_q)\gamma\cdot\varepsilon_{_V}(\lambda_V)\gamma_5
u_P(h)\,\vert\,V,q;\lambda_V,\lambda_q\,
\rangle\nonumber\\
&&+\phi_{_S}\bar{u}(\lambda_q)
u_P(h)\vert\,S,q;\lambda_S,\lambda_q\, \rangle,\label{eqn.19}
\end{eqnarray}
where $\vert\,P,h\, \rangle=\Psi^{\lambda_0}_P(qD)_{_{LF}}$ and
$\vert\,D,q;\lambda_D,\lambda_q\,
\rangle=D^{\lambda_D}(q_1q_2)q_3^{\lambda_q}$ are the Fock states
with $D$ denoting the vector or scalar diquark. Consequently, we can
introduce the effective quark-diquark-nucleon vertex,
$\bar{u}(\lambda_q)\gamma\cdot\varepsilon_{_V}(\lambda_V)\gamma_5
u_P(\lambda_0)$ for the axial vector diquark and
$\phi_{_S}\bar{u}(\lambda_q) u_P(\lambda_0)$ for the scalar diquark.

The $\Delta$(1232) wavefunctions in the SU(6) quark-diquark model
are
\begin{eqnarray}
\Psi_{\Delta^+}^{\pm \frac{1}{2}}\left(qD\right)
&=&\pm\frac{1}{3}\left[2 V^0\left(ud\right)u^{\uparrow,\downarrow}+
\sqrt{2} V^{\pm 1}\left(ud\right)u^{\downarrow,\uparrow} \right.\nonumber\\
&&\quad\quad\left.+ \sqrt{2}
V^0\left(uu\right)d^{\uparrow,\downarrow} + V^{\pm
1}\left(uu\right)d^{\downarrow,\uparrow}
\right],\nonumber\\
\Psi_{\Delta^+}^{\pm \frac{3}{2}}\left(qD\right)
&=&\pm\frac{1}{\sqrt{3}}\left[\sqrt{2}V^{\pm
1}\left(ud\right)u^{\uparrow,\downarrow} + V^{\pm
1}\left(uu\right)d^{\uparrow,\downarrow}\right].\nonumber\\\label{eqn.20}
\end{eqnarray}
After similar treatment, we can take the effective
quark-diquark-delta vertex as
$\bar{u}(\lambda_q)\varepsilon_{_V}^{\mu}(\lambda_V)
u_{\mu}(\lambda_0)$ for the axial vector diquark. Here
$u_{\mu}(\lambda_0)$ is the light-front Rarita-Schwinger spinor
defined in Eq.~(\ref{eqn.5}). In the $\Delta$ case, there is no
coupling with the scalar diquark as shown in Eq.(\ref{eqn.20}).

In the above quark-diquark two-body Fock state basis, the proton and
$\Delta$(1232) spin LF wavefunctions can be expressed by the full
relativistic effective vertices. This covariant description provides
us a convenient tool to treat the diagonal and off-diagonal elements
of the plus current in the lowest Fock state expansion.

\section{Light-front calculation in a solvable covariant model}
\label{sec.4}

 From the above fully relativistic description of the nucleon and
$\Delta$(1232) spin wavefunctions, we can take an effective
treatment of the $N$-$\Delta$ electromagnetic transition from the
covariant field theory. Similar to the description in the vector
meson case~\cite{BCJ02,BCJ03,CJ04,CJ05}, a solvable model based on
the covariant model of ($3+1$)-dimensional field theory enables us
to derive the form factors in LF helicity basis.

\begin{figure*}
\includegraphics[width=6in]{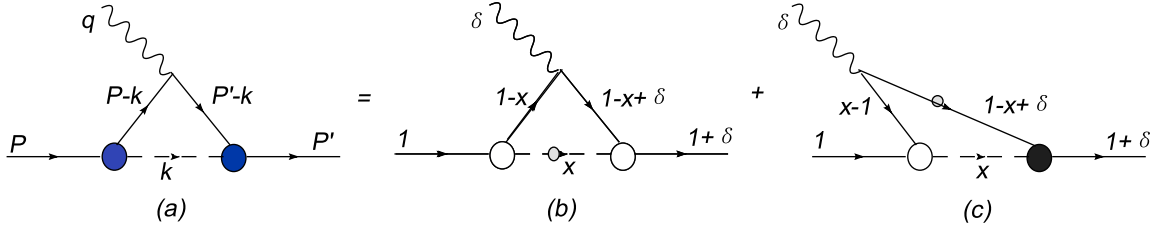}
\caption{The covariant triangle diagram (a) is represented as the
sum of a LF valence diagram (b) defined in the region $0<k^+<p^+$
and the nonvalence diagram (c) defined in $p^+<k^+<p'^+$.
$\delta=q^+/p^+=p'^+/p^+-1$. The white and black blobs at the
baryon-quark vertices in (b) and (c) represent the LF wavefunction
and non-wavefunction vertices, respectively. The small circles in
(b) and (c) represent the (on-shell) mass pole of the quark
propagator determined from the $k^-$-integration. \label{Fig1}}
\end{figure*}

The matrix element $J^\mu_{h'h}(0)$ of the electromagnetic current
with constituents of masses $m_q$ and $m_d$ obtained from the
covariant diagram of Fig.~\ref{Fig1}(a) is given by
%
\begin{eqnarray}
J^\mu_{h'h}(0)&=& iN_cg_\Delta g_P\int\frac{d^4k}{(2\pi)^4}
\frac{S_\Lambda(k-p)S^\mu_{h'h}S_\Lambda(k-p')}{[k^2-m^2_d+i\varepsilon]}\nonumber\\
&&\times
\frac{1}{[(k-p)^2-m^2_q+i\varepsilon][(k-p')^2-m^2_q+i\varepsilon]},\nonumber\\
\label{eqn.21}
\end{eqnarray}
where $g_\Delta$ and $g_P$ are the coupling constants and $N_c$ is
the number of colors. In Eq.(\ref{eqn.21}), $S^\mu_{h'h}$ is the
spinor matrix element
\begin{eqnarray}
S^\mu_{h'h}&=&\bar{u}_{\rho}(p',h')
(\not\!{p}^{\prime}-\not\!k+m)\gamma^\mu(\not\!p-\not\!k+m)\gamma_{\nu}\gamma_5
u(p,h)\nonumber\\
&&\times\sum_{\lambda_V}\epsilon^{\rho}(k,\lambda_V)\epsilon^{*\nu}(k,\lambda_V),\label{eqn.22}
\end{eqnarray}
where $\epsilon^{\rho}(k,\lambda_V)$ is the polarized vector for
axial vector diquark with helicity $\lambda_V$. To regularize the
covariant triangle-loop in ($3+1$) dimension, we replace the
point-like photon-vertex $\gamma^\mu$ by a non-local (smeared)
photon-vertex $S_\Lambda(p'-k)\gamma^\mu S_\Lambda(p-k)$, where
$S_\Lambda(p-k)=\Lambda^2/((k-p)^2-\Lambda^2+i\varepsilon)$ and
$\Lambda$ plays the role of a momentum cut-off similar to the
Pauli-Villars regularization~\cite{BCJ1}.

 From Eqs.~(\ref{eqn.8}) and~(\ref{eqn.21}), we can perform
manifestly covariant calculation in LF helicity basis to obtain the
form factors $G_i(i=1,2,3)$ using dimensional regularization. Here
we list only the essential steps for the derivation of the covariant
form factors: \\
\noindent
$(i)$ We reduce the five propagators into the sum of
three propagators using
\begin{eqnarray}
\frac{1}{D_\Lambda D_0 D_k D'_0 D'_\Lambda}&=&
\frac{1}{(\Lambda^2-m^2_q)^2}\biggl( \frac{1}{D'_\Lambda} - \frac{1}{D'_0}\biggr) \nonumber\\
&&\times \frac{1}{D_k}\biggl(\frac{1}{D_\Lambda} - \frac{1}{D_0}
\biggr),\label{eqn.23}
\end{eqnarray}
where
\begin{eqnarray}
&&D_\Lambda = (k-p)^2-\Lambda^2 + i\ep, \nonumber\\
&&D_0=(k-p)^2-m^2_q+i\ep,\nonumber\\
&&D_k= k^2-m^2_q + i\ep,\label{eqn.24}
\end{eqnarray}
and $D'_{0[\Lambda]}=D_{0[\Lambda]}(p\to p')$.\\
\noindent
$(ii)$ We use the
Feynman parametrization for the three propagators, e.g.,
\begin{eqnarray}
\frac{1}{D_k D_0 D'_0}&=& 2\int^1_0 dx\int^{1-x}_0 dy
\nonumber\\
&&\times\frac{1}{[D_k(1-x-y) + D_0 x + D'_0 y]^3},\label{eqn.25}
\end{eqnarray}
and compute the integrals over the momentum by shifting the
integration variable from $k$ to $k'=k-(xp + yp')$ in both numerator
and denominator which can be written as $D= k'^2+xp^2 +yp'^2-(xp +
yp')^2-(x+y) m^2_q-(1-x-y)m^2_d$. \\
\noindent $(iii)$ We make a Wick rotation of Eq.~(\ref{eqn.21})
in $D$-dimension to regularize the integral.\\
\noindent
Following the above procedures $(i)$ -$(iii)$, we finally obtain the
covariant form factors. The procedure is very similar to
Ref.~\cite{BJ05}, from which one can find some technical details.

In the present work of LF dynamics~(LFD), we shall use only the
plus-component $J^+_{h'h}$ of the current matrix elements
$J^{\mu}_{h'h}$ in the calculation of form factors. One can directly
calculate the term $S^+_{h'h}$ with $k^-=k^-_{\rm pole}$, which
depends on the integration region of $k^+$. However, for the purpose
of a clear understanding of the physics implied in LFD, we separate
$S^+_{h'h}$ into the on-energy-shell part and the off-energy-shell
part by using the following identity
\begin{widetext}
\begin{eqnarray}
\sum_{\lambda_V}\epsilon^{\rho}(k,\lambda_V)\epsilon^{*\nu}(k,\lambda_V)
& =&-g^{\rho\nu}+\frac{k^{\rho}k^{\nu}}{m^2_d}=
-g^{\rho\nu}+\frac{[k^{\rho}_{\rm on}+{g^{\rho +}\over 2}(k^- -
k^-_{\rm on})][k^{\nu}_{\rm on}+{g^{\nu +}\over 2}(k^- - k^-_{\rm
on})]}{m^2_d},\label{eqn.26}
\end{eqnarray}
\end{widetext} where the metric $g^{+-}=2$ and the subscript (on)
denotes the on-energy-shell ($k^2=m^2_q$) quark propagator, i.e.
$k^-=k^-_{\rm on}=(m^2_q+\vec{k}^2_\perp)/k^+$. Then we separate the
term $S^+_{h'h}$ into the on-energy-shell propagating part,
$(S^+_{h'h})_{\rm on}$, and the off-energy-shell part,
$(S^+_{h'h})_{\rm off}$
\begin{eqnarray}
S^+_{h'h}= (S^+_{h'h})_{\rm on} + (S^+_{h'h})_{\rm
off}.\label{eqn.27}
\end{eqnarray}
The off-energy-shell part $(S^+_{h'h})_{\rm off}$ can be divided
into
\begin{eqnarray}
(S^+_{h'h})_{\rm{off}}&=&(k^--k^{-}_{\rm
on})(T^+_{h'h})_{\rm{off}}+(k^--k^{-}_{\rm
on})^2(V^+_{h'h})_{\rm{off}}.\nonumber\\
\label{eqn.29}
\end{eqnarray}
In the Appendix, we present the explicit expressions of
$(S^+_{h'h})_{\rm on}$ and $(S^+_{h'h})_{\rm off}$ and list the
calculated results of each helicity amplitude in DYW frame.

Now, by doing the integration over $k^-$, one can derive the LF
amplitudes from the covariant amplitude in Eq.~(\ref{eqn.21}). In
terms of LF components one gets
\begin{eqnarray}
J^{+}_{h'h}(0) & = & {i N_c g_\Delta g_P  } \int {{\mathrm
d}k^+{\mathrm d}k^- {\mathrm d}\vec{k}_{\perp}^2\over 2 (2 \pi)^4}
 \nonumber\\
&&\times{1\over k^+(k^+ - p^+)^2 (k^+ - p'^+)^2 }
 \nonumber\\
&&\times \frac{1}{(k^- - k'^-_{zm})(k^- - k'^-_{z\Lambda})}\frac{
\Lambda^2 S^+_{h'h}\Lambda^2}{(k^- - k_{vs}^-) }
 \nonumber\\
&&\times \frac{1}{(k^- - k_{zm}^-)(k^- - k'^-_{z\Lambda})}
,\label{eqn.31}
\end{eqnarray}
where the five poles in $k^-$ are explicitly given by
\begin{eqnarray}\left\{
  \begin{array}{l}
    k_{vs}^- = {m_d^2 + \vec{k}_{\perp}^2 \over k^+} - { i\varepsilon\over k^+},  \\
    k_{zm}^- = p^- + {m_q^2 + (\vec{k}_{\perp}-\vec{p}_{\perp})^2 \over k^+ - p^+} - {i \varepsilon \over k^+
-p^+},\\
    k'^{-}_{zm}  =  p'^- +{m_q^2 + (\vec{p'}_{\perp} -
\vec{k}_{\perp})^2 \over k^+ - p'^+} - {i \varepsilon \over k^+
- p'^+},  \\
    k^{-}_{z\Lambda}  =  p^- + {{\Lambda}^2 +
(\vec{k}_{\perp}- \vec{p}_{\perp})^2 \over k^+ - p^+} - {i
\varepsilon \over k^+
-p^+},  \\
    k'^{-}_{z\Lambda}  =  p'^- +{{\Lambda}^2 + (\vec{k}_{\perp}
- \vec{p'}_{\perp})^2 \over k^+ - p'^+} - {i \varepsilon \over k^+ -
p'^+}.
  \end{array}
\right.\label{eqn.32}
\end{eqnarray}

Let us now consider the pole structure of Eq. (\ref{eqn.31}) due to
only the constituent propagators. As is well known, applying the
Cauchy theorem, four different cases should be analyzed: $k^+ < 0$,
$0 \leq k^+ \leq p^+$, $p^+ \leq k^+ \leq p'^+$ and $k^+
> p'^+$. The first and fourth cases do not contribute to the
integral over $k^-$, because all the poles in Eq. (\ref{eqn.31})
have imaginary parts with the same sign. It can be easily seen that
the only surviving contributions come from the regions $0 \leq k^+
\leq p^+$ and $p^+ \leq k^+ \leq p'^+$. In the former region, the
integration over $k^-$ can be done in the lower half-plane, so that
only the pole at $k^- = k_{vs}^-$ (the spectator-pole) contributes
yielding the valence diagram. In the latter region, the contour in
the upper half-plane picks up only the pole at $k^- = k'^-_{zm}$ and
$k^- = k'^-_{z\Lambda}$ yielding the so-called nonvalence diagram.

To avoid the complexity of treating double $k^-$-poles, we
re-decompose the product of five energy denominators in
Eq.~(\ref{eqn.21}) into a sum of four energy denominators:
\begin{eqnarray}\label{eq.17}
\frac{1}{D_\Lambda D_0 D_k D'_0 D'_\Lambda}&=&
\frac{1}{m^2_q-\Lambda^2}\frac{1}{D_k}\frac{1}{D'_0}
\frac{1}{D_0D_\Lambda}\nonumber\\
&&+
\frac{1}{\Lambda^2-m^2_q}\frac{1}{D_k}\frac{1}{D'_\Lambda}\frac{1}{D_0D_\Lambda}.
\label{eqn.33}
\end{eqnarray}
Then the first term has only one single pole $k^- = k'^-_{zm}$ and
the second term has another single pole $k^- = k'^-_{z\Lambda}$ in
the upper half-plane.

Therefore, the covariant diagram shown in Fig.~\ref{Fig1}(a) is in
general equivalent to the sum of the LF valence diagram in
Fig.~\ref{Fig1}(b) and the nonvalence diagram in Fig.~\ref{Fig1}
(c), where $\delta=q^+/p^+=p'^+/p^+-1$. The two LF time-ordered
contributions to the residues correspond to the two poles in $k^-$,
the one coming from the interval (I) $0<k^+<p^+$[the ``valence
contribution"], and the other one from (II) $p^+<k^+<p'^+$ [the
``nonvalence contribution" or the ``zero mode" when $\delta\to
0$].

\subsection{Valence Contribution}
\label{subsec.4.1}

In the region $0<k^+<p^+$ as shown in Fig.~\ref{Fig1}(b), the pole
$k^-=k^-_{\rm on}=(m^2_d + \vec{k}^2_\perp -i\varepsilon)/k^+$
(i.e., the spectator quark), is located in the lower half of the
complex $k^-$-plane.  Thus, the Cauchy integration formula for the
$k^-$-integral in Eq.~(\ref{eqn.31}) gives in this region the
following result for the plus current;
\begin{eqnarray}
J^+_{h'h}(V) &=&N_c\int \frac{{\mathrm d}x{\mathrm
d}^2\vec{k}_{\perp} }{2(2\pi)^3} {S^+_{h'h}(V)\over
x(1-x')^2(1-x)^2}\nonumber\\
&&\times{g_\Delta \Lambda^2\over (M_{\Delta}^2-{\cal
M'}^2_0)(M_{\Delta}^2-{\cal M'}^2_\Lambda)}\nonumber\\
&&\times {g_P \Lambda^2\over (M_{P}^2-{\cal M}^2_0)(M_{P}^2-{\cal
M}^2_\Lambda)},\label{eqn.34}
\end{eqnarray}
where $x'={x\over 1+\delta}$, the invariant masses of the initial
state are
\begin{eqnarray}
{\cal M}_0^2&=&{\vec{k}_{\perp}^2+m_d^2\over
x}+{\vec{k}_{\perp}^2+m_q^2\over
1-x},\nonumber\\
{\cal M}_\Lambda^2&=&{\vec{k}_{\perp}^2+m_d^2\over
x}+{\vec{k}_{\perp}^2+\Lambda^2\over 1-x},\label{eqn.35}
\end{eqnarray}
and the invariant masses of the final state are
\begin{eqnarray}
{\cal M'}_0^2&=&{\vec{k'}_\perp^2+m_d^2 \over
x'}+{\vec{k'}_\perp^2+m_q^2\over
1-x'},\nonumber\\
{\cal M'}_\Lambda^2&=&{\vec{k'}_{\perp}^2+m_d^2 \over
x'}+{\vec{k'}_{\perp}^2+\Lambda^2\over 1-x'}.\label{eqn.36}
\end{eqnarray}
Here, $\vec{k'}_\perp=\vec{k}_{\perp}-x'\vec{q}_{\perp}$ is the
Drell-Yan assignment~\cite{DYW}.

Due to $\gamma^+\gamma^+=0$ and the pole $k^-_{\rm on}$, both quark
and diquark are on-energy-shell. As one can easily see from
Eq.(\ref{eqn.29}), $S^{+val}_{h'h}=(S^{+}_{h'h})_{\rm on}$.
Therefore, the physical interpretation of the LF plus current matrix
element given by Eq.~(\ref{eqn.34}) is manifest in terms of LF wave
functions, i.e., a convolution of the initial and final state LF
wavefunctions. Such identification is not possible for the covariant
calculation.
 From Eq.~(\ref{eqn.34}),
the initial and the final state LF wave
functions are the smeared LF wavefunctions given by
\begin{eqnarray}
\varphi_\Delta(x',\vec{k'}_\perp)&=&{g_\Delta \Lambda^2\over
(M_{\Delta}^2-{\cal M'}^2_0)(M_{\Delta}^2-{\cal
M'}^2_\Lambda)},\nonumber\\
\varphi_P(x,\vec{k}_\perp)&=&{g_P \Lambda^2\over (M_{P}^2-{\cal
M}^2_0)(M_{P}^2-{\cal M}^2_\Lambda)}.\label{eqn.37}
\end{eqnarray}

\subsection{Zero-mode contribution}
\label{subsec.4.2}

In the region $p^+<k^+<p'^+(=p^++q^+)$ as shown in
Fig.~\ref{Fig1}(c), the poles are at
$k^-=p'^-+{(\vec{k}_{\perp}-\vec{p'}_\perp)^2+m_q^2-i\epsilon\over
k^+-p'^+}$ (from the struck quark propagator) and
$k^-=p'^-+{(\vec{k}_{\perp}-\vec{p'}_\perp)^2+\Lambda^2-i\epsilon\over
k^+-p'^+}$ (from the smeared quark-photon vertex $S_\Lambda(k-p')$).
Both of them are located in the upper half plane of the complex
$k^-$ space.

When we do the Cauchy integration over $k^-$ to obtain the LF
time-ordered diagrams, we decompose the product of five energy
denominators into a sum of two terms in Eq.~(\ref{eqn.33}), i.e.,
\begin{eqnarray}
J^+_{h'h}(Z)
&=&J^+_{h'h}(Z,k'^-_{zm})+J^+_{h'h}(Z,k'^-_{z\Lambda}).\label{eqn.38}
\end{eqnarray}

After above decomposition, the first term in Eq.~(\ref{eqn.38})
that has only one pole $k^-=k'^-_{zm}$ is given by
\begin{eqnarray}
J^+_{h'h}(Z,k'^-_{zm}) &=&{N_c\, \Lambda^4 \over(\Lambda^2-m^2_q)}
\int_{1}^{1+\delta} \frac{dxd^2\vec{k}_{\perp} }{2(2\pi)^3}\nonumber\\
&&\times{ g_\Delta\over (M_{\Delta}^2-{\cal
M'}_0^2)}{S^+_{h'h}(Z,k^-=k'^-_{zm})\over x'(x-1)^2(x'-1)}\nonumber\\
&&\times{g_P\over ({q}^2-{\cal M'}^2_{qm})({q}^2-{\cal
M'}^2_{q\Lambda})},\label{eqn.39}
\end{eqnarray}
where ${q}^2={\vec q}^2_\perp+M_{\Delta}^2-M_{P}^2$
and the invariant masses are defined as
\begin{eqnarray}
{\cal M'}_{qm}^2&=&{(\vec{k}_{\perp}-\vec{q}_{\perp})^2+m_q^2 \over
1-x'}+{\vec{k}_{\perp}^2+m_q^2\over
(x-1)/(1+\delta)},\nonumber\\
{\cal M'}_{q\Lambda}^2&=&{(\vec{k}_{\perp}-\vec{q}_{\perp})^2+m_q^2
\over 1-x'}+{\vec{k}_{\perp}^2+\Lambda^2\over
(x-1)/(1+\delta)}.\label{eqn.40}
\end{eqnarray}
Here, the initial state wavefunction is the photon smeared LF wave
function
\begin{eqnarray}
\varphi_q(x,\vec{k}_\perp;q^2)&=&{\Lambda^2\over ({q}^2-{\cal
M'}^2_{qm})({q}^2-{\cal M'}^2_{q\Lambda})}.\label{eqn.41}
\end{eqnarray}
The second term in Eq.~(\ref{eqn.38}) that also has a single pole
$k^-=k'^-_{z\Lambda}$ is given by
\begin{eqnarray}
J^+_{h'h}(Z,k'^-_{z\Lambda}) &=&{N_c\, \Lambda^4
\over(\Lambda^2-m^2_q)} \int_{1}^{1+\delta}
\frac{dxd^2\vec{k}_{\perp} }{2(2\pi)^3}\nonumber\\
&&\times{ g_\Delta\over (M_{\Delta}^2-{\cal
M'}_0^2)}{S^+_{h'h}(Z,k^-=k'^-_{z\Lambda})\over
x'(x-1)^2(1-x')}\nonumber\\
&&\times{g_P\over ({q}^2-{\cal M'}^2_{\Lambda m})({q}^2-{\cal
M'}^2_{\Lambda\Lambda})},\label{eqn.42}
\end{eqnarray}
where the invariant masses are
\begin{eqnarray}
{\cal M'}_{\Lambda
m}^2&=&{(\vec{k}_{\perp}-\vec{q}_{\perp})^2+\Lambda^2 \over
1-x'}+{\vec{k}_{\perp}^2+m_q^2\over
(x-1)/(1+\delta)},\nonumber\\
{\cal
M'}_{\Lambda\Lambda}^2&=&{(\vec{k}_{\perp}-\vec{q}_{\perp})^2+\Lambda^2
\over 1-x'}+{\vec{k}_{\perp}^2+\Lambda^2\over
(x-1)/(1+\delta)}.\label{eqn.43}
\end{eqnarray}
Adding Eqs.~(\ref{eqn.39}) and~(\ref{eqn.42}), we get the full
result in nonvalence region. In general, the terms in $S^+_{h'h}$
include the on-energy-shell and off-energy-shell contributions in
the nonvalence region; {\it e.g.},  $S^+_{h'h}(k^-)=(S^+_{h'h})_{\rm
on} + (S^+_{h'h})_{\rm off}(k^-)$. In addition, although the
nonvalence part of transition form factors can not be directly
written as the convolution of the conventional wavefunctions like
that of the valence part, we can still utilize the LF Bethe-Salpeter
approach and relate the nonvalence part to the valence
part~\cite{JC01}. This allows the convolution formalism also for the
nonvalence part. Once both valence and nonvalence parts can be
expressed as the convolution formulism, we can replace these smeared
LF wavefunctions by the conventional LFQM Gaussian wavefunctions and
recover the usual LFQM description.

In the $q^+\to 0$ limit, the integration range of the nonvalence
region, $p^+<k^+<p'^+(=p^++q^+)$, shrinks to zero so that the
nonvalence contribution reduces to the zero mode
contribution~\cite{BCJ02,BCJ03,CJ04,CJ05}
\begin{equation}
(J^+_{h'h})_{\rm z.m.} = \lim_{\delta\to 0}(J^+_{h'h})_{\rm
nv}=\lim_{\delta\to 0}\int^{1+\delta}_1 dx(\cdots).\label{eqn.44}
\end{equation}
The non-vanishing zero-mode contribution occurs only if the
integrand $(\cdots)$ in Eq.~(\ref{eqn.44}) behaves as $\sim k^-({\rm
i.e.}(1-x)^{-1})$. Note that there is no zero-mode contribution in
the case that the integrand is $k^-$-independent or behaves like
$k^-(k^+-p^+)^n(n\geq 1)$. For the plus current, the zero-mode
contribution comes from the helicity matrix elements of the
propagators $S^+_{h'h}(k^-)$, specifically only from the
instantaneous part, and neither from the on-energy-shell propagating
part nor the energy denominator. In the Appendix, we analyze the
zero mode contribution for each helicity element by using the power
counting method~\cite{CJ05}.

Performing the calculation in Eq.~(\ref{eqn.39}), the result for
$J^+_{h'h}(Z,k'^-_{zm})$ is given by
\begin{eqnarray}
J^+_{h'h}(Z,k'^-_{zm})
&=&{N_c\,g_\Delta\,g_P\,\Lambda^4\over(\Lambda^2-m^2_q)}
\int_{1}^{1+\delta} \frac{dxd^2\vec{k}_{\perp} }{2(2\pi)^3}{1\over
{\cal B}_{m}{\cal
B}_{\Lambda}}\nonumber\\
&&\times\Big[{(x-1-\delta)\over x} T^+_{h'h}+{1\over x^2}{\cal
A}V^+_{h'h}\Big],\nonumber\\\label{eqn.45}
\end{eqnarray}
where
\begin{eqnarray}
{\cal A}&=&{M_{\Delta}^2+\vec{q}_{\perp}^2 \over
1+\delta}(x-1-\delta)x+[(\vec{k}_{\perp}-\vec{q}_{\perp})^2+m_q^2]x\nonumber\\
&&-[\vec{k}_{\perp}^2+m_d^2](x-1-\delta),
\end{eqnarray}
and
\begin{eqnarray}
{\cal B}_{m}&=&({M_{\Delta}^2+\vec{q}_{\perp}^2\over
1+\delta}-M_{P}^2)(x-1)(x-1-\delta)
\nonumber\\
&&+[(\vec{k}_{\perp}-\vec{q}_{\perp})^2+m_q^2](x-1)\nonumber\\
&&-[\vec{k}_{\perp}^2+m_q^2](x-1-\delta),\nonumber\\
{\cal B}_{\Lambda}&=&({M_{\Delta}^2+\vec{q}_{\perp}^2\over
1+\delta}-M_{P}^2)(x-1)(x-1-\delta)
\nonumber\\
&&+[(\vec{k}_{\perp}-\vec{q}_{\perp})^2+m_q^2](x-1)\nonumber\\
&&-[\vec{k}_{\perp}^2+\Lambda^2](x-1-\delta).\label{eqn.46}
\end{eqnarray}
Similarly, we can get the result for $J^+_{h'h}(Z,k'^-_{z\Lambda})$.
If we write $x=1+\delta y$ and ${\mathrm d}x=\delta {\mathrm d}y$ in
Eq.~(\ref{eqn.45}) and $J^+_{h'h}(Z,k'^-_{z\Lambda})$, then the
integral over $y$ runs from 0 to 1 as $x$ runs from 1 to $1+\delta$.
Taking $q^+\to 0$ limit ($\delta\to 0$), the zero mode contribution
is given by
\begin{widetext}
\begin{eqnarray}
(J^+_{h'h}(Z))_{\rm z.m.} &=&{N_c \,g_\Delta \,g_p
\Lambda^4\over(\Lambda^2-m^2_q)} \int_{0}^{1} \frac{ {\mathrm
d}y{\mathrm d}^2\vec{k}_{\perp}
}{2(2\pi)^3} \nonumber\\
&&\times\left\{{-(U^+_{h'h})_{\rm
off}[(\vec{k}_{\perp}-\vec{q}_{\perp})^2+m_q^2]\,y-(T^+_{h'h})_{\rm
off}(1-y)\over \left[ ((\vec{k}_{\perp}-\vec{q}_{\perp})^2+m_q^2)
y+(\vec{k}_{\perp}^2+m_q^2) (1-y)\right]\left[
((\vec{k}_{\perp}-\vec{q}_{\perp})^2+m_q^2)
y+(\vec{k}_{\perp}^2+\Lambda^2) (1-y)\right]}\right.\nonumber\\
&&\left.+{(U^+_{h'h})_{\rm
off}[(\vec{k}_{\perp}-\vec{q}_{\perp})^2+\Lambda^2]\,y+(T^+_{h'h})_{\rm
off}(1-y)\over \left[
((\vec{k}_{\perp}-\vec{q}_{\perp})^2+\Lambda^2)
y+(\vec{k}_{\perp}^2+\Lambda^2) (1-y)\right]\left[
((\vec{k}_{\perp}-\vec{q}_{\perp})^2+\Lambda^2)
y+(\vec{k}_{\perp}^2+m_q^2) (1-y)\right]}\right\}.\label{eqn.47}
\end{eqnarray}
\end{widetext}
where ${(U^+_{h'h})}_{ \rm off}$ corresponds to ${(V^+_{h'h})_{ \rm
off}\over 1-x}$. As shown in the Appendix, $(J^+_{{1\over 2},{1\over
2}}(Z))_{\rm z.m.}$ and $(J^+_{{1\over 2},-{1\over 2}}(Z))_{\rm
z.m.}$ do not vanish while the others ($(J^+_{{3\over 2},{1\over
2}}(Z))_{\rm z.m.}$ and $(J^+_{{3\over 2},-{1\over 2}}(Z))_{\rm
z.m.}$) do.
\section{Numerical results}
\label{sec.5}

\begin{figure}
\includegraphics{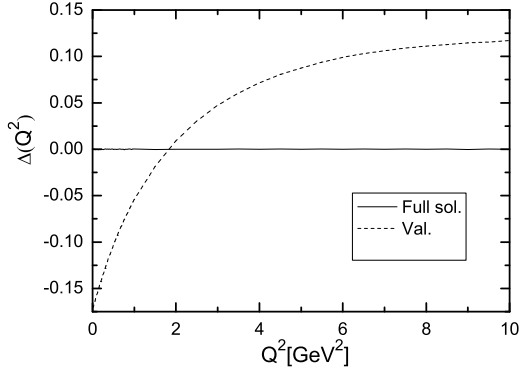}
 \caption{The
solid curve represents the angular conditions $\Delta(Q^2)$
including the zero mode while the dash curve is the angular
condition excluding the zero mode. $\Delta(Q^2)$ equals to zero
exactly after considering zero mode.}\label{fig.dt}
\end{figure}

\begin{figure}
\includegraphics{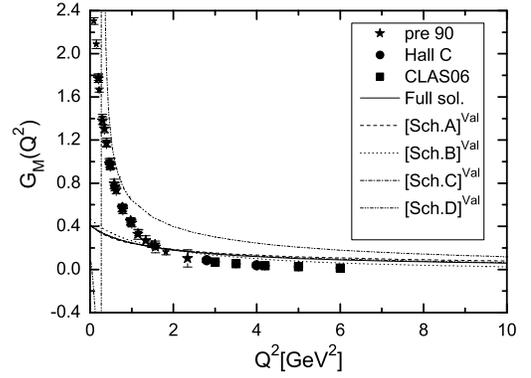}
\caption{The magnetic form factor $G_M(Q^2)$. The filled circles
corespond to JLab Hall C data~\cite{bib:frolov}. The filled squares
 are from CLAS experiment~\cite{bib:clas06}. The filled stars are
 from pre-90 data at DESY, Bonn and
 SLAC~\cite{bib:desy68,bib:desy72,bib:desy722,bib:bonn74,bib:slac75}.
 The solid curve represents the full solutions of
the form factor $G_M(Q^2)$ in four schemes which have the same
curve. The dash curve represents the valence part of the form
factors in scheme A, the dot curve in scheme B, the dash-dot curve
in scheme C and the dash-dot-dot curve in scheme D.
}\label{fig.gmdata}
\end{figure}

\begin{figure}
\includegraphics{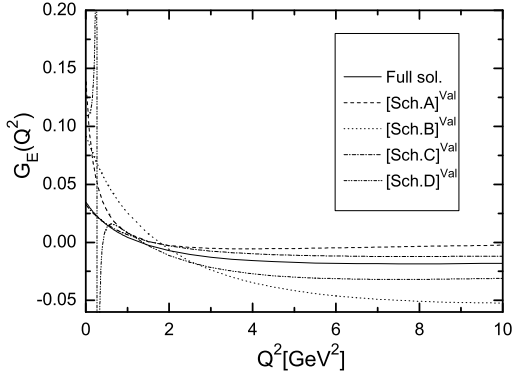}
\caption{The electric form factor $G_E(Q^2)$ in three different
schemes with and without zero mode. The five curves correspond to
the same cases as in Fig.~\ref{fig.gmdata}.}\label{fig.ge}
\end{figure}

\begin{figure}
\includegraphics{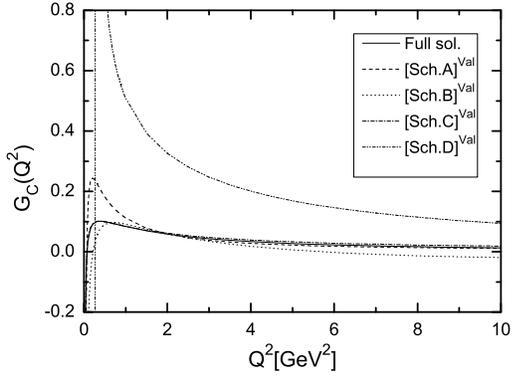}
\caption{The Coulomb form factor $G_C(Q^2)$ in three different
schemes with and without zero mode.  The five curves correspond to
the same cases as in Fig.~\ref{fig.gmdata}.}\label{fig.gc}
\end{figure}

\begin{figure}
\includegraphics{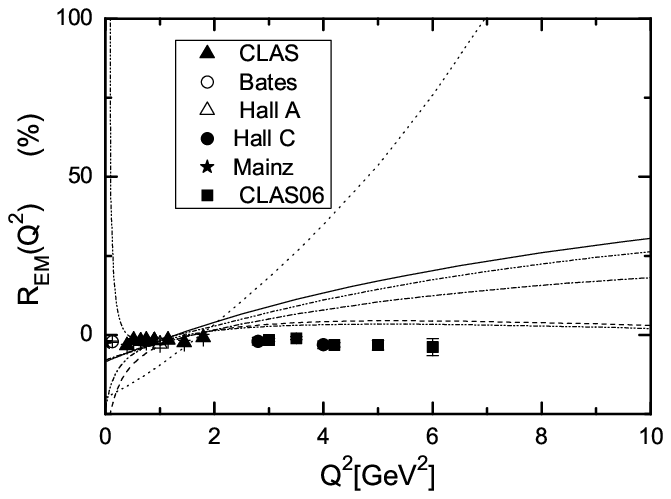}
\caption{The ratios $R_{EM}$.
           Both the filled squares~\cite{bib:clas06} and
           filled  triangles~\cite{bib:colejoo} are the CLAS results.
           The open triangles are from an
           JLab Hall A  experiment~\cite{bib:hall-A}.
           The filled  circles: JLab Hall-C~\cite{bib:frolov}. Open circles: Bates~\cite{bib:bates}.
           Filled stars: Mainz~\cite{bib:mami}.  The five curves correspond to
the same cases as in Fig.~\ref{fig.gmdata}. }\label{fig.remdata}
\end{figure}

\begin{figure}
\includegraphics{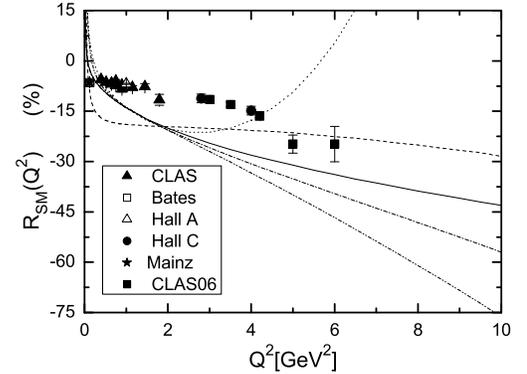}
\caption{The ratios $R_{SM}$.
           Both the filled squares~\cite{bib:clas06} and
           filled  triangles~\cite{bib:colejoo} are the CLAS results.
           The open triangles are from an
           JLab Hall A  experiment~\cite{bib:hall-A}.
           The filled  circles: JLab Hall-C~\cite{bib:frolov}. Open circles: Bates~\cite{bib:bates}.
           Filled stars: Mainz~\cite{bib:mami2}.  The five curves correspond to
the same cases as in Fig.~\ref{fig.gmdata}. }\label{fig.rsmdata}
\end{figure}

In this section, we present the numerical results for the transition
form factors and verify that all of the form factors obtained in the
LF helicity basis are in complete agreement with the manifestly
covariant results. However, we do not aim at finding the best-fit
parameters to describe the experimental data of the $N$-$\Delta$
transition properties. Rather, we simply take typical parameters
used before. Nevertheless, our model calculations have a generic
structure and the essential findings from our calculations may apply
to the more realistic models. The quantitative results would
certainly depend on the details of the model.

In our numerical calculations, we thus use the quark and diquark
masses as $m_q=0.55$ GeV and $m_d=0.70$ GeV, respectively, noting
that a rather large value of quark mass is allowed to handle the
$\Delta$ particle and take $\Lambda=1.8$ GeV as in the calculation
of spin-one meson form factors~\cite{BCJ02}. Also, we simply take
the values of parameters $g_P$ and $g_\Delta$ as typical strong
coupling constant 1. In this work, we do not consider the flavor
structure of nucleon and $\Delta$(1232) to show that the zero-mode
issue is completely independent from whether the model is
``spectator diquark" or ``struck diquark". In addition, we make the
taxonomical decompositions of the full results into the valence and
nonvalence contributions to facilitate a quantitative comparison
between the full results and the valence parts of the form factors
($G_M,G_C,G_E$) in four different schemes. To show the importance of
zero mode, we also compare the angular condition $\Delta(Q^2)$
including the zero mode with the one without the zero mode.

In Fig.~\ref{fig.dt}, we show the angular condition $\Delta(Q^2)$
given by Eq.~(\ref{eqn.9}), neglecting the zero-mode contribution.
From the dashed curve in the Fig.~\ref{fig.dt}, we find that
$\Delta(Q^2)\neq 0$ and thus the angular condition is violated. If
we include the zero-mode contribution, $\Delta(Q^2)$ is exactly zero
in all four schemes as shown by the solid curve. Thus, the zero mode
contribution is crucial to get the correct full solution.

The $N$-$\Delta$ transition is given predominantly in the magnetic dipole
(M1) type. In the instant-form quark model picture, the $N$-$\Delta$
transition is described by a spin flip of a quark in the s-wave
state, which leads to the magnetic dipole (M1) type transition. In
LFQM, because of the Melosh transformation, the relativistic effect
 is also included. We show the magnetic form factor $G_M$ in
Fig.~\ref{fig.gmdata}. The full solutions (solid curve) are obtained
from four different prescriptions and they all turn out to give
exactly the same result as the covariant one. To estimate the zero
mode contribution, we also plot the valence contribution for each
prescription, i.e., the dashed curve for scheme A, dotted curve for
scheme B, dash-dotted curve for scheme C and dash-dot-dotted curve
for scheme D, respectively. In particular, the valence contribution
of scheme D yields a singular behavior near $Q^2 = (M_\Delta -
M_N)M_N$ due to the denominator structure $K_D$ in
Eq.(\ref{eqn.13}). However, the full solution of scheme D including
the zero mode contribution renders the regular result depicted by
the solid curve coinciding with the full solutions of other schemes
(A,B and C) as expected. Because we simply take the typical
parameters, and consider neither the flavor structure nor the
struck-diquark contribution, it is not really surprising to see a
rather large disagreement of our present model calculation from the
experimental data as shown in Fig.~\ref{fig.gmdata}. The LF
wavefunction we took in the present model calculation is also not
realistic since it is just coming from the smearing procedure of the
triangle loop. The purpose of our present analysis is not to
reproduce the data, but to show the necessity of taking into account
the zero mode contribution to get the correct result without any
scheme dependence. Our analysis here also indicates that the
zero-mode issue is completely independent from whether the model is
``spectator diquark" or ``struck diquark". In another words, the
Lorentz covariance must work independently for each contribution
whether the struck constituent is quark or diquark.

The electric and Coulomb form factors $G_E(Q^2)$ and $G_C(Q^2)$ are
shown in Figs.~\ref{fig.ge} and ~\ref{fig.gc}, respectively. As
anticipated, the full solution in each prescription is exactly the
same as the covariant results. However, unlike the case of $G_M$ the
valence contributions of $G_E(Q^2)$ and $G_C(Q^2)$ to the full
solution are quite different even in the schemes of A, B and C
although the scheme D has already been expected to be different due
to the denominator structure $K_D$ in Eq.(\ref{eqn.13}). This
suggests that $G_C$ and $G_E$ are more sensitive to the zero mode
than $G_M$. A very interesting point is that the values of both
$G_E$ and $G_C$ are not zero at $Q^2=0$ although the s-wave
instant-form SU(6) quark model predicts that the values of both
$G_E$ and $G_C$ are exactly zero in non-relativistic
limit~\cite{BM65}. The starting point of our model is also the
s-wave instant-form SU(6) quark wavefunction. On the other hand, in
LFQM the relativistic effect and orbital angular
momentum~\cite{Bro02} are responsible for the electric (E2) and
Coulomb (C2) quadrupole transitions. It has been shown that there
are indeed different mechanisms to produce non-zero values of $G_E$
and $G_C$. Even in the non-relativistic quark model (excluding the
s-wave SU(6) breaking model), the d-wave admixture in the nucleon
and $\Delta$ wave functions allows for the electric (E2) and Coulomb
(C2) quadrupole transitions.

These physical form factors $G_M(Q^2)$, $G_E(Q^2)$ and $G_C(Q^2)$
can be related to other physical quantities $R_{EM}(Q^2)$ and
$R_{SM}(Q^2)$. The relations are as follows~\cite{Jones,DEK}
\begin{eqnarray}
R_{EM}(Q^2) & = & \frac{E2(Q^{2})}{M1(Q^{2})} =
\frac{E_{1+}}{M_{1+}} = - \frac{G_E(Q^2)}{G_M(Q^2)},
 \nonumber \\
R_{SM}(Q^2) & = & \frac{C2(Q^{2})}{M1(Q^{2})} =
\frac{S_{1+}}{M_{1+}} = -\frac{Q_+Q_-}{4 M_{\Delta}^2}
\frac{G_C(Q^2)}{G_M(Q^2)},\nonumber\\\label{eqn.48}
\end{eqnarray}
where $Q_{\pm}=\sqrt{(M_\Delta\pm M_P)^2+Q^2}$. In
Fig.~\ref{fig.remdata}, we show the physical quantity $R_{EM}$
obtained by using the light-front helicity basis both from full
solutions and from only valence part of physical form factors in
four (A,B,C,D) schemes. The corresponding results for the physical
quantity $R_{SM}$ are shown in Fig.~\ref{fig.rsmdata}. In
Fig.~\ref{fig.remdata}, the calculated value of $R_{EM}$ at $Q^2=0$
is $-8.33\%$. In non-relativistic quark models with unbroken
SU(6)-spin-flavour symmetry, however, $R_{EM}$ is predicted to be
exactly zero~\cite{BM65}, whereas a broken SU(6) symmetry yields the
values ranging from 0 to $-0.2\%$~\cite{Isgur82}. It is also known
well that the perturbative QCD (pQCD) power counting rule predicts
that $R_{EM}$ approaches to 1~\cite{carl} and $R_{SM}$ approaches to
constant up to logarithmic corrections~\cite{XJi04} in the limit
$Q^2 \rightarrow \infty$. The 12 GeV upgrade of JLab facility is
anticipated to shed some light on the applicability of PQCD at large
$Q^2$. For the better agreement between our model calculation and
the experimental data of $G_M$, $R_{EM}$ and $R_{SM}$, we need to
consider the flavor structure and use the more realistic Gaussian LF
wavefunctions.

\section{Conclusion}
\label{sec.6} In this work, we investigated the electromagnetic
transition between nucleon and $\Delta$(1232) using both the
manifestly covariant technique in the helicity basis and the
light-front (LF) calculation for the matrix elements of the plus
component current. In the LF calculation, the full solution is
decomposed into the valence and zero-mode contribution. Using the
power counting method and exact calculation for each helicity
amplitude, we found that only two helicity amplitudes receive the
zero mode contribution. From the numerical computation, we have
explicitly shown that the full results of the LF calculation
including the zero mode contribution are in complete agreement with
the covariant one without prescription dependence. Our analysis also
showed that the zero-mode issue is completely independent from
whether the model is ``spectator diquark" or ``struck diquark" and
the Lorentz covariance must work independently for each contribution
whether the struck constituent is quark or diquark. Because our
model calculations have a generic structure, i.e., a convolution of
initial and final state LF wavefunctions, our calculations can apply
to more realistic cases. The treatment for the zero mode
contribution can also be extended to other nucleon-resonance
transition processes.

\acknowledgments

This work is partially supported by National Natural Science
Foundation of China (Nos.~10421503, 10575003, 10528510), by the Key
Grant Project of Chinese Ministry of Education (No.~305001), and by
the Research Fund for the Doctoral Program of Higher Education
(China). CRJ's work was supported by a grant from the U.S.
Department of Energy under Contract No. DE-FG02-03ER41260.

\begin{widetext}
\appendix*
\section{Summary of spinor terms in helicity amplitude}
\label{sec.A1}

The on-energy-shell part $(S^+_{h'h})_{\rm on}$ is given by
$(S^+_{h'h})_{\rm on}=(S^+_{1h'h})_{\rm on}+(S^+_{2h'h})_{\rm on}$,
where
\begin{eqnarray}
(S^+_{1h'h})_{\rm{on}}&=&-\bar{u}_{\rho}(p',h')
(\not\!{p}^{\prime}-\not\!k+m)\gamma^+(\not\!p-\not\!k+m)\gamma^{\rho}\gamma_5
u(p,h),\nonumber\\
(S^+_{2h'h})_{\rm{on}}&=&{1\over
m_d^2}\bar{u}_{\rho}(p',h')\,k^{\rho}_{\rm on}
(\not\!{p}^{\prime}-\not\!k+m)\gamma^+(\not\!p-\not\!k+m)\not\!k_{\rm
on}\gamma_5 u(p,h),\label{eqA.1}
\end{eqnarray}
and the off-energy-shell part $(S^+_{h'h})_{\rm off}$ is given by
\begin{eqnarray}
(S^+_{h'h})_{\rm{off}}&=&(k^--k^{-}_{\rm
on})(T^+_{h'h})_{\rm{off}}+(k^--k^{-}_{\rm
on})^2(V^+_{h'h})_{\rm{off}}\nonumber\\
&=&(k^--k^{-}_{\rm on})(T^+_{1h'h})_{\rm{off}}+(k^--k^{-}_{\rm
on})(T^+_{2h'h})_{\rm{off}}+(k^--k^{-}_{\rm
on})^2(V^+_{h'h})_{\rm{off}},\label{eqA.2}
\end{eqnarray}
where
\begin{eqnarray}
(T^+_{1h'h})_{\rm{off}}&=&{1\over m_d^2}\bar{u}_{-}(p',h')
(\not\!{p}^{\prime}-\not\!k+m)\gamma^+(\not\!p-\not\!k+m)\not\!k_{\rm
on}\gamma_5 u(p,h),\nonumber\\
(T^+_{2h'h})_{\rm{off}}&=&{1\over
m_d^2}\bar{u}_{\rho}(p',h')\,k^{\rho}_{\rm on}
(\not\!{p}^{\prime}-\not\!k+m)\gamma^+(\not\!p-\not\!k+m)\gamma_-\gamma_5 u(p,h),\nonumber\\
(V^+_{h'h})_{\rm{off}}&=&{1\over m_d^2}\bar{u}_{-}(p',h')
(\not\!{p}^{\prime}-\not\!k+m)\gamma^+(\not\!p-\not\!k+m)\gamma_-\gamma_5
u(p,h).\label{eqA.3}
\end{eqnarray}

In DYW frame with the LF gauge, the explicit forms of the spinor
terms in Eqs.~(\ref{eqA.1}) and~(\ref{eqA.2}) for each helicity
component are given as follows.

(I) helicity (${3\over 2},{1\over 2}$)-component:
\begin{eqnarray}
(S^+_{{3\over 2},{1\over 2}})_{\rm{on}}
&=&[k^L-xq^L]\left\{x^2M_{P}m_q[m_q+(1-x)M_{\Delta}]+xm_d^2[m_q-(1-x)M_{\Delta}+2(1-x)M_{P}]\right.\nonumber\\
&&\left.-k^Lk^R[(1-x)M_{\Delta}+(1-x)m_q-xM_{P}]-x^2q^Lk^R(m_q+M_{P})\right\}\nonumber\\
&&-[k^L+xq^L]m_d^2[m_q+(1-x)M_{\Delta}],\label{eqA.4}
\end{eqnarray}
where $k^{R(L)}=k^1\pm{\mathrm i}k^2$ and
\begin{eqnarray}
(T^+_{1{3\over 2},{1\over 2}})_{\rm{off}}&=&0,\nonumber\\
(T^+_{2{3\over 2},{1\over 2}})_{\rm{off}} &=&{2\sqrt{2}(p^+)^2\over
m_d^2}(1-x)
(k^L-xq^L)\left[m_q+(1-x)M_{\Delta}\right],\nonumber\\
(V^+_{{3\over 2},{1\over 2}})_{\rm{off}}&=&0.\label{eqA.5}
\end{eqnarray}
In the nonvalence region where $k^-\sim 1/(1-x)$, we find that all
the off-shell terms are regular as $q^+\to 0$ (or equivalently $x\to
1$). Therefore, there is no zero-mode in the helicity (${3\over
2},{1\over 2}$) component.

(II) helicity (${3\over 2},-{1\over 2}$)-component:
\begin{eqnarray}
(S^+_{{3\over 2},-{1\over 2}})_{\rm{on}}
&=&[k^L-xq^L]\left\{k^Lk^R(k^L-xq^L)+(1-x)m_d^2(k^L+xq^L)+xm_q(x^2M_{P}q^L+m_qk^L)\right.\nonumber\\
&&\left.+x(1-x)k^L(M_{P}M_{\Delta}+m_qM_{P}+m_qM_{\Delta})\right\},\label{eqA.6}
\end{eqnarray}
and
\begin{eqnarray}
(T^+_{1{3\over 2},-{1\over 2}})_{\rm{off}}&=&0,\nonumber\\
 (T^+_{2{3\over 2},-{1\over 2}})_{\rm{off}}
&=&{-2\sqrt{2}(p^+)^2\over m_d^2}(1-x)(k^L-xq^L)^2,\nonumber\\
(V^+_{{3\over 2},-{1\over 2}})_{\rm{off}}&=&0.\label{eqA.7}
\end{eqnarray}
Again, the off-shell terms are regular as $x\to 1$. Therefore, there
is no zero-mode in the helicity $({3\over 2},-{1\over 2})$
component.

(III) helicity (${1\over 2},{1\over 2}$)-component:
\begin{eqnarray}
(S^+_{1{1\over 2},{1\over 2}})_{\rm{on}}&=&4\sqrt{{2\over
3}}{p^+\over
M_{\Delta}}\left\{[m_qM_{P}+(1-x)M_{\Delta}^2-q^Lq^R][m_q+(1-x)M_{\Delta}]\right.\nonumber\\
&&\left.+[M_{P}k^R+(1-x)(M_{P}-M_{\Delta})q^R][k^L-xq^L]\right\}\nonumber\\
&&-4\sqrt{{2\over
3}}p^+\left\{[(1-x)q^L+k^L][k^R-xq^R]+[m_q+(1-x)M_{P}][m_q+(1-x)M_{\Delta}]\right\},\label{eqA.8}
\end{eqnarray}
\begin{eqnarray}
(S^+_{2{1\over 2},{1\over 2}})_{\rm{on}}&=&-4\sqrt{{1\over
6}}{p^+\over m_d^2x^2}
[m_d^2-x^2M_{\Delta}^2+(k^R-xq^R)(k^L-xq^L)]\nonumber\\
&&\times \left\{[m_q+(1-x)M_{\Delta}][x^2M_{P}m_q-(1-x)m_d^2-k^Lk^R]+(k^L-xq^L)k^Rx(m_q+M_{P})\right\}\nonumber\\
&&-4\sqrt{{1\over 6}}{p^+ \over
m_d^2x}(k^L-xq^L)\left\{(-x^2m_qM_{P}+(1-x)m_d^2+k^Lk^R)(k^R-q^Rx)
\right.\nonumber\\
&&+\left. k^Rx(M_{P}+m_q)(m_q+(1-x)M_{\Delta}
)\right\},\label{eqA.9}
\end{eqnarray}
and
\begin{eqnarray}
(T^+_{1{1\over 2},{1\over 2}})_{\rm{off}}
&=&-2\sqrt{\frac{2}{3}}{(p^+)^2 \over
xm_d^2M_{\Delta}}\left\{[m_q+(1-x)M_{\Delta}][x^2M_{P}m_q-(1-x)m_d^2]\right.\nonumber\\
&&\left.+k^Lk^R[xM_{P}-(1-x)m_q-(1-x)M_{\Delta}]-k^Rq^Lx^2(m_q+M_{P})\right\},\nonumber\\
 (T^+_{2{1\over 2},{1\over 2}})_{\rm{off}}
&=& 2\sqrt{\frac{2}{3}}{(p^+)^2\over
xm_d^2M_{\Delta}}(1-x)[m_q+(1-x)M_{\Delta}][m_d^2-x^2M_{\Delta}^2+(k^L-xq^L)(k^R-xq^R)]
\nonumber\\
&&-2\sqrt{\frac{2}{3}}{(p^+)^2\over m_d^2}(1-x)(k^L-xq^L)(k^R-xq^R), \nonumber\\
(V^+_{{1\over 2},{1\over 2}})_{\rm{off}}
&=&2\sqrt{\frac{2}{3}}{(p^+)^3 \over m_d^2M_{\Delta}}
(1-x)[m_q+(1-x)M_{\Delta}].\label{eqA.10}
\end{eqnarray}
 From the power counting method of the longitudinal momentum
fraction, one can see that both $(T^+_{1{1\over 2},{1\over 2}})_{\rm
off}$ and $(V^+_{{1\over 2},{1\over 2}})_{\rm off}$ yield a singular
behavior in the integrand of Eq.(\ref{eqn.45}) giving effectively
the zero-mode contribution to the helicity (${1\over 2},{1\over 2}$)
component.

(IV) helicity $({1\over 2},-{1\over 2})$-component:
\begin{eqnarray}
(S^+_{1{1\over 2},-{1\over 2}})_{\rm{on}}&=&4\sqrt{{2\over
3}}{p^+\over
M_{\Delta}}\left\{q^L(M_{P}+m_q)[m_q+(1-x)M_{\Delta}]\right.\nonumber\\
&&\left.+
[(k^R+(1-x)q^R)q^L+(1-x)M_{P}M_{\Delta}-(1-x)M_{\Delta}^2][k^L-xq^L]\right\}\nonumber\\
&&-4\sqrt{{2\over
3}}{p^+}(1-x)[m_q+(1-x)M_{\Delta}]q^{L},\label{eqA.11}
\end{eqnarray}
\begin{eqnarray}
(S_{2{\frac{1}{2}},-{1\over 2}}^{+})_{\mathrm{on}} &=&-4p^{+}\sqrt{{\frac{1%
}{6}}}{\frac{1}{x^{2}m_{d}^{2}}}[m_{d}^{2}-x^{2}M_{\Delta
}^{2}+(k^{R}-xq^{R})(k^{L}-xq^{L})]  \notag \\
&&\times \left\{
(-x^{2}m_{q}M_{P}+(1-x)m_{d}^{2}+k^{L}k^{R})(k^{L}-q^{L}x)+k^{L}x(M_{P}+m_{q})(m_{q}+(1-x)M_{\Delta })\right\}   \notag \\
&&+4p^{+}\sqrt{{\frac{1}{6}}}\frac{1}{xm_d^2}(k^{L}-xq^{L})\left\{
[m_{q}++(1-x)M_{\Delta
}][x^{2}M_{P}m_{q}-(1-x)m_{d}^{2}-k^{L}k^{R}]\right.
\notag \\
&&\left. +(k^{R}-xq^{R})xk^{L}(m_{q}+M_{P})\right\} ,\label{eqA.12}
\end{eqnarray}
and
\begin{eqnarray}
(T^+_{1{1\over 2},-{1\over 2}})_{\rm{off}}
&=&-2\sqrt{\frac{2}{3}}{(p^+)^2 \over
xm_d^2M_{\Delta}}\left\{q^Lx[x^2m_qM_{P}-(1-x)m_d^2-k^Lk^R]\right.\nonumber\\
&&\left.+k^L[(1-x)(m_d^2+xm_qM_{P})+(1+\delta)(k^Lk^R+m_q^2)+x(1-x)M_{\Delta}(m_q+M_{P})]\right\},
\nonumber\\
 (T^+_{2{1\over 2},-{1\over 2}})_{\rm{off}}
&=& 2\sqrt{\frac{2}{3}}{(p^+)^2\over
xm_d^2M_{\Delta}}(1-x)(k^L-xq^L)[m_d^2-x^2M_{\Delta}^2+k^L-xq^L)(k^R-xq^R)]
\nonumber\\
&&-2\sqrt{\frac{2}{3}}{(p^+)^2\over m_d^2}
(1-x)(k^L-xq^L)[m_q+(1-x)M_{\Delta}], \nonumber\\
(V^+_{{1\over 2},-{1\over 2}})_{\rm{off}}
&=&-2\sqrt{\frac{2}{3}}{(p^+)^3 \over
m_d^2M_{\Delta}}(1-x)(k^L-xq^L).\label{eqA.13}
\end{eqnarray}
Again from the power counting method, $(T^+_{1{1\over 2},-{1\over
2}})_{\rm off}$ and $(V^+_{{1\over 2},-{1\over 2}})_{\rm off}$ cause
a singular behavior in the integrand of Eq.(\ref{eqn.45}) giving
again effectively the zero-mode contribution to the helicity
(${1\over 2},-{1\over 2}$) component.
\end{widetext}

\def\Journal#1#2#3#4{{#1} {\bf#2}, #3 (#4)}
\def\NCA{\rm Nuovo Cimento}
\def\NPA{{\rm Nucl. Phys.} A}
\def\NPB{{\rm Nucl. Phys.} B}
\def\PLB{{\rm Phys. Lett.}  B}
\def\PRT{\rm Phys. Rep.}
\def\PRL{\rm Phys. Rev. Lett.}
\def\PRD{{\rm Phys. Rev.} D}
\def\PRC{{\rm Phys. Rev.} C}
\def\ZPC{{\rm Z. Phys.} C}
\def\AP{{\rm Ann. Phys.} }


\begin{thebibliography}{99}

\bibitem{Bro98}S.~J.~Brodsky, H.-C.~Pauli and S.~S.~Pinsky,
\Journal{\PRT}{301}{299}{1998}; S. J. Brodsky, T. Huang, and G. P.
Lepage, in Particles and Fields-2, edited by A. Z. Capri and A. N.
Kamal (Plenum, New York, 1983).


\bibitem{BCJ02} B. L. G. Bakker, H.-M. Choi, and C.-R. Ji,
\Journal{\PRD}{65}{116001}{2002}.

\bibitem{BCJ03} B. L. G. Bakker, H.-M. Choi, and C.-R. Ji,
\Journal{\PRD}{67}{113007}{2003}.

\bibitem{CJ04} H.-M. Choi and C.-R. Ji,
\Journal{\PRD}{70}{053015}{2004}.

\bibitem{CJ05} H.-M. Choi and C.-R. Ji,
\Journal{\PRD}{72}{013004}{2005}.


\bibitem{GK} I. L. Grach and L. A. Kondratyuk,
Sov. J. Nucl. Phys. {\bf 39}, 198 (1984).

\bibitem{Weber90}
H.~J.~Weber, \Journal{\AP}{207}{417}{1991}.

\bibitem{Simula95}
F.~Cardarelli, E.~Pace, G.~Salm\'e and S.~Simula,
\Journal{\PLB}{371}{7}{1996}.

\bibitem{Melosh}
E. Wigner, Ann. Math. {\bf 40}, 149 (1939); H.~J.~Melosh,
\Journal{\PRD}{9}{1095}{1974}; L.  A. Kondratyuk and M. V.
Terent'ev, Yad. Fiz.  {\bf 31}, 1087 (1980) [Sov. J. Nucl. Phys.
{\bf 31}, 561 (1980)].

\bibitem{Jones}
H.~F.~Jones and M.~D.~Scadron, \Journal{\AP}{81}{1}{1973}.

\bibitem{DEK}
R.~C.~E.~Devenish, T.~S.~Eisenschitz and J.~G.~Korner,
\Journal{\PRD}{14}{3063}{1976}.

\bibitem{DYW}
S.~D.~Drell and T.~M.~Yan, \Journal{\PRL}{24}{181}{1970}; G. West,
\Journal{\PRL}{24}{1206}{1970}.

\bibitem{Ang}
C.~E.~Carlson and C.-R.~Ji, \Journal{\PRD}{67}{116002}{2003}.

\bibitem{Ma96}
B.-Q.~Ma, \Journal{\PLB}{375}{320}{1996}.

\bibitem{Bro02}
S. Brodsky, D. S. Hwang, B.-Q. Ma and I. Schmidt,
\Journal{\NPB}{593}{311}{2001}.

\bibitem{Qing02}
B.-Q.~Ma, D.~Qing, and I.Schmidt, \Journal{\PRC}{65}{035205}{2002}.

\bibitem{Diquark76}
M.~I.~Pavkovi\'c, \Journal{\PRD}{13}{2128}{1976}.

\bibitem{Rolnick}
W. B. Rolnick, \Journal{\PRD}{25}{2439}{1982}.

\bibitem{KW}
W.~Konen and H.~J.~Weber, \Journal{\PRD}{41}{2201}{1990}.

\bibitem{Weber98}
M. Beyer, C. Kuhrts, H.J. Weber, \Journal{\AP}{269}{129}{1998}.

\bibitem{BCJ1}
B.~L.~G.~Bakker, H.-M.~Choi, and C.-R.~Ji,
\Journal{\PRD}{63}{074014}{2001}.

\bibitem{BJ05}
B.~L.~G.~Bakker and C.-R.~Ji, \Journal{\PRD}{71}{053005}{2005}.

\bibitem{JC01} C.-R. Ji and H.-M. Choi,
\Journal{\PLB}{513}{330}{2001}.

\bibitem {bib:frolov}
V. V. Frolov {\it et al.}, \Journal{\PRL}{82}{45}{1999}.

\bibitem{bib:clas06}
M.~Ungaro {\it et al.}, \Journal{\PRL}{97}{112003}{2006}.

\bibitem{bib:desy68}
W. Bartel {\it et al.}, \Journal{\PLB}{28}{148}{1968}.

\bibitem{bib:desy72}
J. C. Alder {\it et al.}, \Journal{\NPB}{46}{573}{1972}.

\bibitem{bib:desy722}
S. Galster {\it et al.}, \Journal{\PRD}{5}{519}{1972}.

\bibitem{bib:bonn74}
K. Batzner {\it et al.}, \Journal{\NPB}{76}{1}{1974}.

\bibitem{bib:slac75}
S. Stein {\it et al.}, \Journal{\PRD}{12}{1884}{1975}.

\bibitem {bib:colejoo}
K. Joo {\it et al.}, \Journal{\PRL}{88}{122001}{2002}.

\bibitem {bib:hall-A}
J.J. Kelly {\it et al.}, \Journal{\PRL}{95}{102001}{2005}.

\bibitem {bib:bates}
C. Mertz {\it et al.}, \Journal{\PRL}{86}{2963}{2001}.

\bibitem {bib:mami}
R. Beck {\it et al.}, \Journal{\PRC}{61}{035204}{2000}.

\bibitem{bib:mami2}
Th.~Pospischil {\it et al.}, \Journal{\PRL}{86}{2959}{2001}.

\bibitem{BM65}
C. Becchi and G. Morpurgo, \Journal{\PLB}{16}{352}{1965}.

\bibitem{Isgur82}
N. Isgur, G. Karl and R. Koniuk, \Journal{\PRD}{25}{2394}{1982}.

\bibitem{carl}
C. E. Carlson, \Journal{\PRD}{34}{2704}{1986}.

\bibitem{XJi04}
A. Idilbi, X. Ji and J.-P. Ma, \Journal{\PRD}{69}{014006}{2004}.


\end{thebibliography}
\end{document}